\documentclass[preprint,showpacs,preprintnumbers,amsmath,amssymb]{revtex4}

\usepackage{color}
\usepackage{graphicx}
\usepackage{dcolumn}
\usepackage{bm}

\def\Vec#1{\mbox{\boldmath $#1$}}

\begin{document}


\title{Quantum Monte-Carlo study of magnetic ordering in ZnV$_2$O$_4$}


\author{Yasuyuki Kato}
\affiliation{Teoretical division and Center for Nonlinear Studies, Los Alamos National Lab, Los Alamos NM 87545 USA}

\date{\today}
\begin{abstract}
	We study the magnetic ordering of Vanadium spinels by Quantum Monte Carlo simulations of a three-band Hubbard model.
	Vanadium spinels, AV$_2$O$_4$, exhibit  a unique ``up-up-down-down'' spin ordering at low temperatures.
	While this magnetic ordering  was originally measured in 1973, its origin has remained unclear for many years due to the lack of unbiased approaches for solving the relevant model.
	A three-band Hubbard model on the spinel lattice (corner sharing tetrahedra) is a minimal Hamiltonian for describing the $t_{2g}$ electrons of the V$^{2+}$ ions.
	One of the main difficulties is that this family of compounds belongs to the elusive intermediate-coupling regime ($U \gtrsim t$) for which there is no small parameter that can justify a perturbative
	expansion. We present a controlled Quantum Monte-Carlo approach to the three-band Hubbard model relevant for this materials that reproduces the up-up-down-down spin ordering.
	The method is free of  the sign problem that is usually the main limiting factor for simulating fermionic systems in dimension higher than one.
\end{abstract}

\maketitle

\section{Introduction}
Interacting electrons in crystals exhibit a  variety of cooperative phenomena. This is particularly evident for Mott insulators whose valence electrons are localized due
to a rather strong intra-orbital Coulomb repulsion. The degree of electronic localization is controlled by the ratio $U/t$, where $t$ is the dominant transfer integral between
different atomic orbitals. The strong  ($U/t \gg 1$) and weak-localization  ($t/U\ll1$) regimes  can be treated by  expanding in the small parameters
$t/U$ or $U/t$. Unfortunately,  this is not true for the intermediate-coupling regime ($U/t \gtrsim 1$) because of the lack of small parameter.
Therefore, it is very important to develop controlled and unbiased  numerical techniques that can address this elusive regime.
The main obstacle for Monte Carlo simulations is the sign problem that can arise from the fermionic statistics  of the electronic degrees of freedom or the frustrated nature of the 
Hamitonian terms. 

\begin{figure}[htp]
\begin{center}
\includegraphics[width=16cm,trim=0 340 0 0, clip ] {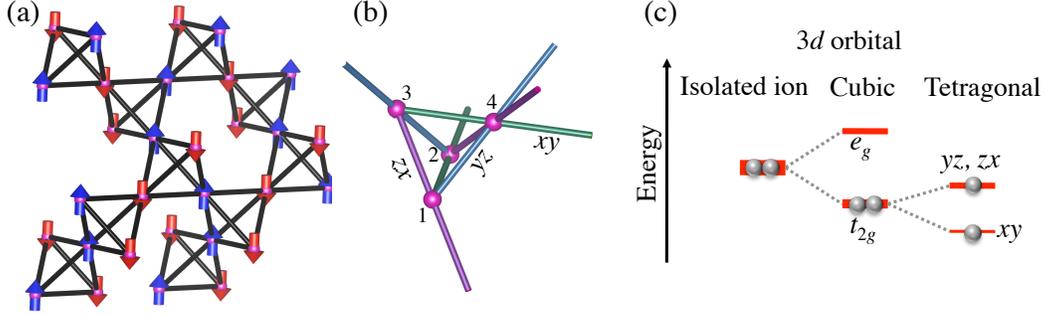}
\end{center}
\caption{
	(a) up-up-down-down magnetic ordering at low temperature, (b) unit cell of pyrochlore  lattice, (c) crystal field splitting due to structural transition.
} \label{fig:znv2o4}
\end{figure}

In this paper, we focus on the Vanadium spinels A$^{2+}$V$^{3+}_2$O$^{2-}_4$, a class of  highly frustrated Mott insulators that belong to the intermediate-coupling regime.
These systems comprise a pyrochlore lattice (corner sharing tetrahedra) of  V$^{3+}$ ions (see Fig. \ref{fig:znv2o4}(a)). 
Each  V$^{3+}$ ion has two in 3-$d$ electrons in the $t_{2g}$ orbitals, while the  O$^{2-}$ and Zn$^{2+}$ ions are in closed shell configurations.
Consequently, the low-energy electronic spectrum consists of spin and orbital excitations that arise from the localized 3-$d$ electrons in the V$^{3+}$ ions.

The lattice symmetry of  AV$_2$O$_4$ is cubic ($Fd{\bar 3}$m) at high temperatures.
The  crystal field produced by the octahedral coordination of  O$^{2-}$ ions splits the  V$^{3+}$ $3d$-orbitals  into  high-energy $e_g$ and low-energy $t_{2g}$ orbitals. 
ZnV$_2$O$_4$ undergoes a structural cubic to tetragonal ($I4_1/amd$) transition at $T=45\sim51$(K) \cite{ueda1997,vasiliev2006}.
This structural transition causes further crystal field splitting of the $t_{2g}$ orbitals  into $xy$-orbital and $yz,zx$-orbitals. 
(See Fig.\ref{fig:znv2o4}(c).)
A magnetic transition to {\it up-up-down-down} magnetic ordering (uudd-MO) occurs at a lower temperature  $T_N=31 \sim 40$(K)  \cite{ueda1997,vasiliev2006}.
The magnetic ordering has a period of 4 lattice sites 
($\cdots\uparrow\uparrow\downarrow\downarrow\cdots$) 
for chains oriented along the $yz$ $(0,\pm1,\pm1)$ and $zx$ $(\pm1, 0,\pm1)$ directions. 
The origin of the  uudd-MO and the lack of orbital ordering in this material have been an open this magnetic ordering was originally reported in 1973 \cite{niziol1973}.
More recent measurements of the pressure dependence of $T_N$ in different spinel vanadates \cite{Blanco-Canosa07} indicate that ZnV$_2$O$_4$ belongs to the intermediate-coupling regime:
$T_N$ decreases with pressure instead of increasing according to the Bloch's law \cite{Bloch66}  that is expected for the strong-coupling regime. 
Previous attempts at explaining these properties were based on strong-coupling  \cite{tsunetsugu2003,motome2004,tchernyshyov2004,maitra2007} or weak-coupling expansions  \cite{chern2011}  whose validity is not guaranteed for
the intermediate-coupling regime relevant for ZnV$_2$O$_4$.

\section{Three-band Hubbard model on the pyrochlore lattice}
The three-band Hubbard model relevant for the family of spinel vandates AV$_2$O$_4$ is 
\begin{eqnarray}
\mathcal{H}_{\rm 3BH} &=&
-\sum_{\alpha,\sigma}\left[ 
t_{\alpha} 
\sum_{ \left\langle {\Vec r}, {\Vec r}' \right\rangle \in \alpha} 
\left( d^{\dag}_{\alpha \sigma {\Vec r}}d_{\alpha \sigma {\Vec r}'} +{\rm H.c.} \right)
\right]
-\sum_{\alpha} \left[
\mu_{\alpha} \sum_{{\Vec r},\sigma} n_{\alpha\sigma{\Vec r}} 
\right]
\nonumber\\
&&+ U\sum_{{\Vec r}, \alpha}\left[
 n_{\alpha \uparrow {\Vec r}}n_{\alpha \downarrow {\Vec r}}
 \right]
+ \left(\frac{U}{2} -\frac{5J}{4} \right) \sum_{{\Vec r} \alpha, \beta\neq\alpha, \sigma, \sigma'}
\left[ 
n_{\alpha \sigma {\Vec r}}n_{\beta \sigma' {\Vec r}}
\right]
\nonumber\\
&&-\frac{J}{4}  \sum_{{\Vec r} \alpha, \beta\neq\alpha}
\left[
 {\vec \sigma}_{\alpha {\Vec r}} \cdot {\vec \sigma}_{\beta {\Vec r}}
 \right]
+J  \sum_{{\Vec r} \alpha, \beta\neq\alpha \sigma} 
\left[ 
	d^{\dag}_{\alpha  \uparrow {\Vec r}}d^{\dag}_{\alpha  \downarrow{\Vec r}} d_{\beta \downarrow{\Vec r}}d_{\beta \uparrow{\Vec r} } 
	+{\rm H.c.}
\right],\label{eq:3BH}
\end{eqnarray}
where the operator $d^{\dag}_{\alpha \sigma {\Vec r}}$ an electron of spin  $\sigma$ in the orbital $\alpha$ of the V$^{3+}$ ion with coordinates ${\Vec r}$,
$n_{\alpha \sigma {\Vec r}}=d^{\dag}_{\alpha \sigma {\Vec r}}d_{\alpha \sigma {\Vec r}}$, and
$\sigma^{\delta}_{\alpha {\Vec r}}=\sum_{\mu,\nu}d^{\dag}_{\alpha \mu {\Vec r}}\sigma^{\delta}_{\mu\nu}d_{\alpha \nu {\Vec r}}$,
where $\sigma^{\delta}$ are the  Pauli matrices.
$t_\alpha$ is the transfer integral between $\alpha$-orbitals of nearest-neighbor V$^{3+}$ ions, $U$ is the intra-orbital Coulomb repulsion, $J$ is Hund's coupling, $\mu_\alpha$ is the  chemical potential
whose orbital dependence is a consequence of the  crystal field splitting induced by the structural transition, $\mu_{yz} = \mu_{zx}=\mu_{xy}-\Delta$, and
$\sum_{\langle {\Vec r},{\Vec r}' \rangle \in \alpha}$ represents summation over all pairs of the nearest-neighbor sites connected by $\alpha$-type bonds.
(As shown in Fig.\ref{fig:znv2o4}(b), the pyrochlore lattice has three types of bonds: $xy$, $yz$, and $zx$.)
This implies that electrons occupying an $\alpha$-orbital can only move along the single chain of $\alpha$-type bonds that contains such orbital. 
Note that the on-site interaction term has only two independent parameters, $U$ and $J$, because of the rotational symmetry of the single-ion Hamiltonian \cite{dagotto2001}.

By expanding around the strong-coupling limit of $\mathcal{H}_{\rm 3BH} $, Tsunetsugu and Motome derived 
an  effective  Kugel-Komuskii Hamiltonian whose  ground state exhibits anti-ferro layered orbital ordering (OO) \cite{tsunetsugu2003}.
Moreover, by adding an  antiferromagnetic exchange between third-nearest-neighbor spins, 
they showed that an uudd-MO is stabilized and coexists with the OO \cite{motome2004}.
Since the OO has not been observed in any member of the AV$_2$O$_4$ family, 
Tchernyshyov and Maitra-Valent{\'\i} proposed alternative orbital orderings  \cite{tchernyshyov2004,maitra2007}. 
On the other hand, by starting from the opposite  weak-coupling limit, Chern and Batista derived  uudd-MO without any orbital ordering 
by using a mean-field theory \cite{chern2011}.
In this paper, we show that our unbiased approach not only explains the observed magnetic ordering but also the lack of the OO 
in the  intermediate coupling-regime relevant for the vanadium spinels.

Quantum Monte-Carlo methods for the simulation of fermionic Hamiltonians such as $\mathcal{H}_{\rm 3BH}$ typically 
suffer from the well-known negative sign problem introduced by the fermionic statistics and by certain  off-diagonal terms.
To avoid the negative sign problem we replace Heisenberg-like Hund's interaction  (fifth term in Eq. (\ref{eq:3BH})) by  an Ising-like term because.
In doing so, we are assuming that the optimal magnetic ordering of $\mathcal{H}_{\rm 3BH}$ is always collinear.
In addition, we eliminate the hopping of singlet pairs between different orbitals of the same atom  (sixth term in Eq. (\ref{eq:3BH})) by assuming that
the probability of double occupancy of a given orbital remains rather low in the intermediate-coupling regime.
The resulting  Hamiltonian,
\begin{eqnarray}
\mathcal{H} &=&
-\sum_{\alpha,\sigma}\left[ 
t_{\alpha} 
\sum_{ \left\langle {\Vec r}, {\Vec r}' \right\rangle \in \alpha} 
\left( d^{\dag}_{\alpha \sigma {\Vec r}}d_{\alpha \sigma {\Vec r}'} +{\rm H.c.} \right)
\right]
-\sum_{\alpha} \left[
\mu_{\alpha} \sum_{{\Vec r},\sigma} n_{\alpha\sigma{\Vec r}} 
\right]
\nonumber\\
&&+ U\sum_{\Vec r} \left[ 
n_{\alpha \uparrow {\Vec r}}n_{\alpha \downarrow {\Vec r}}
\right]
+ \frac{U-2J}{2}  \sum_{{\Vec r} \alpha, \beta\neq\alpha, \sigma, \sigma'} 
\left[ 
	n_{\alpha \sigma {\Vec r}}n_{\beta \sigma' {\Vec r}}
\right]
\nonumber\\
&&-\frac{J}{2}  \sum_{{\Vec r} \alpha, \beta\neq\alpha, \sigma}
\left[ 
	n_{\alpha \sigma {\Vec r}}  n_{\beta \sigma {\Vec r}} 
\right],
\label{eq:model}
\end{eqnarray}
has only diagonal on-site interactions.
Since all electrons are restricted to move along one-dimensional chains, there is no fermionic sign problem  (fermions cannot be exchanged) and the only remaining challenge for efficient simulations is the 
the presence of geometric frustration.


\section{Method}
We apply the  world-line quantum Monte-Carlo method (QMC) \cite{kawashima2004}  to the Hamiltonian (\ref{eq:model}).
We use a modified directed-loop algorithm for the updating procedure  \cite{kato2009}. 
In this algorithm, we insert a pair of discontinuities to the world-line configuration in ($d+1$)-dimensional space, 
and move one of them stochastically by updating the world-line configuration.
When the  discontinuity comes back to the creation point where the other discontinuity is located, 
 these discontinuities annihilate each other. 
Although the Hamiltonian (\ref{eq:model}) does not cause any negative sign problem, 
the geometrically frustrated nature of the underlying lattice generates a ``freezing problem'' at low temperatures.
To solve this problem, we introduce both thermal and quantum annealing in our QMC simulations.
In the thermal annealing process, we start from an inverse temperature $\beta_{\rm init}$ and add $\delta\beta$ every $N_{\rm ann}$ Monte-Carlo sweeps.
We choose $\beta_{\rm init}=1.0 ({\rm eV}^{-1})$, $N_{\rm ann}=300$, and $\delta\beta =(\beta-\beta_{\rm init})/(N_{\rm change}-1)$ so that the inverse temperature is $\beta$
after $N_{\rm change}$ temperature changes.
Because of the rather strong interaction between electrons, the discontinuity is often geometrically trapped and localized. 
To avoid this ``freezing of the simulation'', we apply a quantum annealing technique.
We add a new term,
\begin{eqnarray}
\mathcal{H}_q &=& -\frac{q}{2} \sum_{{\Vec r},\alpha, \beta\neq\alpha, \sigma,\sigma'}
\left[
	\left(
	d^{\dag}_{\alpha\sigma{\Vec r}}	d^{\dag}_{\beta\sigma'{\Vec r}}
	+d^{\dag}_{\alpha\sigma{\Vec r}}	d_{\beta\sigma'{\Vec r}}
	+d^{\dag}_{\alpha\sigma{\Vec r}}	
	 d^{\dag}_{\beta\sigma'{\Vec r}}d_{\beta{\bar{\sigma'}}{\Vec r}}
	+{\rm H.c.}
	\right)
\right. 
\nonumber\\
&&\left.
	+d^{\dag}_{\alpha\sigma{\Vec r}}	d_{\alpha {\bar\sigma}{\Vec r}}	
	d^{\dag}_{\beta\sigma'{\Vec r}}	d_{\beta {\bar\sigma'}{\Vec r}}	
\right],
\end{eqnarray}
to $\mathcal{H}$ to liberate the trapped discontinuity.
We use the re-weighting technique to calculate the physical quantities of ${\mathcal H}$ 
from the simulation with ${\mathcal H}+{\mathcal H}_q$.
The procedure is rather simple.
We sample the world-line configurations only without any vertices corresponding to ${\mathcal H}_q$.  
We choose $q$ as $q\beta=0.002 \ll 1$ so that we can sample ~90\% of world line configurations.


We define the unit cell of the pyrochlore lattice as shown in Fig. \ref{fig:znv2o4}(c).
The primitive vectors are
	${\Vec a}_1 = (1/2,1/2,0)$,
 	${\Vec a}_2 =(1/2,0,1/2)$, and
 	${\Vec a}_3 =(0,1/2,1/2)$.
Each unit cell is labeled by ${\Vec R}=\sum_{i}n_i{\Vec a}_i$ ($n_i$ is integer).
The coordinates of the four ions in the unit cell are
	${\Vec e}_1 = (1/4,0,0)$,
 	${\Vec e}_2 =( 0,1/4,0)$,
 	${\Vec e}_3 =(0,0,1/4)$, and
	${\Vec e}_4 =(1/4,1/4,1/4)$.
To reveal the existence of uudd-MO and estimate the magnetic transition temperature, we calculate the susceptibility
\begin{eqnarray} 
\chi_{s} = \frac{1}{N_{\rm site}}\left\langle
\left(\sum_{\Vec R} L^{(1)}_{\Vec R}\right)^2
+
\left(\sum_{\Vec R} L^{(2)}_{\Vec R}\right)^2
\right\rangle,
\end{eqnarray}
where 
\begin{eqnarray} 
L^{(1)}_{\Vec R}&\equiv&  \sum_{\alpha} \left( 
	{\sigma}^{z}_{\alpha {\Vec R}+{\Vec e}_1} 
	-{\sigma}^{z}_{\alpha {\Vec R}+{\Vec e}_2} 
	+{\sigma}^{z}_{\alpha {\Vec R}+{\Vec e}_3} 
	-{\sigma}^{z}_{\alpha {\Vec R}+{\Vec e}_4} 
	\right), \nonumber\\
L^{(2)}_{\Vec R}&\equiv&  \sum_{\alpha} \left( 
	{\sigma}^{z}_{\alpha {\Vec R}+{\Vec e}_1} 
	-{\sigma}^{z}_{\alpha {\Vec R}+{\Vec e}_2} 
	-{\sigma}^{z}_{\alpha {\Vec R}+{\Vec e}_3} 
	+{\sigma}^{z}_{\alpha {\Vec R}+{\Vec e}_4} 
	\right), \nonumber
\end{eqnarray}
$L^{(1,2)}_{\Vec R}$ is the order parameter for uudd-MO, i.e.,
$\langle L^{(1)}_{\Vec R}\rangle$ or $\langle L^{(2)}_{\Vec R}\rangle$ is finite in uudd-MO.

\section{Results}
\begin{figure}[htp]
\begin{center}
\includegraphics[width=14cm,trim=0 300 0 0, clip ] {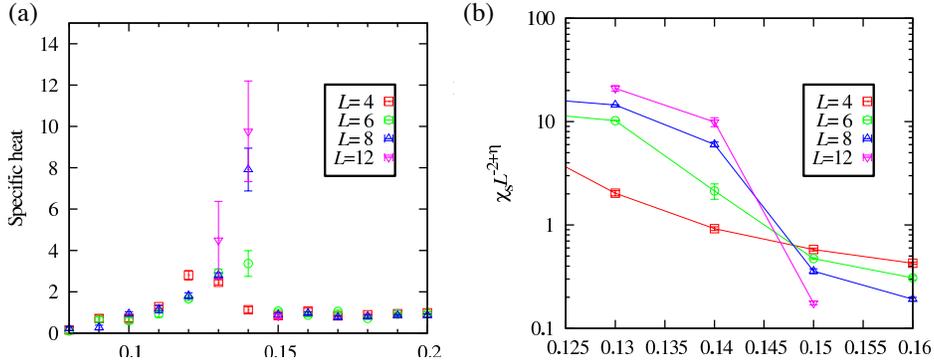}
\end{center}
\caption{
	Results of quantum Monte Carlo simulation with 
	$U=3.9$ (eV), 
	$J=0.8$ (eV), 
	$\Delta = 1$ (eV), 
	$\mu_{xy} = 3.05$ (eV),
	$t=0.8$ (eV), (a) specific heat , (b) finite-size scaling of $\chi_s$ by assuming second order phase transition
	with 3D $Z_2$ symmetry breaking (i.e., $\eta$  =0.04). 
}
\label{fig:qmc}
\end{figure}

\begin{figure}[htp]
\begin{center}
\includegraphics[width=8cm,trim=0 0 0 0, clip ] {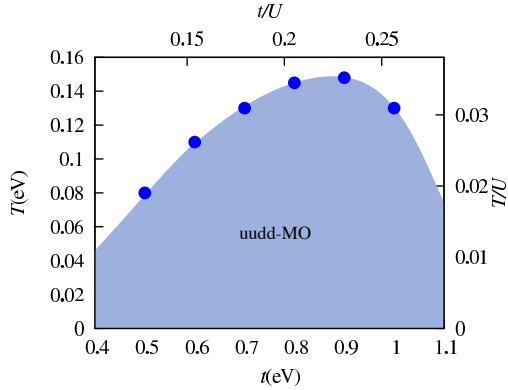}
\end{center}
\caption{
	Phase diagram at $U=3.9$ (eV), $J=0.8$ (eV) , $\Delta=1$ (eV), and $\mu_{xy}=3.05$ (eV)
	with $t=t_{xy}=t_{yz}=t_{zx}$.
	Blue shaded regime is guide to line.
} \label{fig:phasediagram}
\end{figure}

We use $U=3.9$ (eV), and $J=0.8$ (eV) close to the previous theoretical works \cite{tsunetsugu2003,maitra2007,pardo2008}.
($t$ is estimated as 0.35(eV) \cite{takubo2006}.)
Since we are focusing on the magnetic order transition, we assume that the crystal has  tetragonal symmetry: $\Delta>0$.
For concreteness, we use $\mu_{xy}=3.05$ (eV) and $\Delta=1.0$ (eV) 
so that one electron occupies $xy$ orbital and the other electron occupies the $yz$ or $zx$ orbitals at low temperatures.
Figure \ref{fig:qmc} shows the results of our QMC simulations at $t=0.8$ (eV).
The finite size scaling of $\chi_sL^{-2+\eta}$ has a crossing point at $T=0.145\pm0.005$ (eV)
where the specific heat curves have a peak.
Figure \ref{fig:phasediagram} shows five transition temperatures estimated by using the same finite-size scaling analysis.

\section{Discussions and Conclusions}

The transition temperature changes non-monotonically as a function of $t$.
By   calculating 
\begin{eqnarray}
\chi_o =\frac{1}{N_{\rm site}} \left\langle
\left[
\sum_{\Vec R} \left(
\Delta n_{{\Vec R}+{\Vec e}_1}
+\Delta n_{{\Vec R}+{\Vec e}_2}
-\Delta n_{{\Vec R}+{\Vec e}_3}
-\Delta n_{{\Vec R}+{\Vec e}_4}
\right)
\right]^2
\right\rangle,
\end{eqnarray}
where $\Delta n_{\Vec r}=\sum_{\sigma}n_{yz\sigma{\Vec r}}-n_{zx\sigma{\Vec r}}$,
we have also confirmed the OO proposed by  Tsunetsugu-Motome type does not appear for the range of parameters 
shown in Figure \ref{fig:phasediagram}.
This result is qualitatively consistent with the above mentioned  experimental observations. On the other hand, the magnitude of $T_N$
is much higher than the observed values, but it drops very rapidly for smaller values of $t/U$. The natural question is if the OO  is still absent
for values of $t/U$ such that $T_N$ becomes comparable to the experimental values. An answer to this question requires of very low-temperature simulations for which
the freezing problem becomes more challenging. It is important to note that the rather high value of 
$\Delta$ that we used for our simulations tends to increase $T_N$ because of the localization of a single electron ($S=1/2$) in the $xy$-orbital. 
We are also  ignoring the $\pi$-bond hopping of electrons that can reduce further the magnitude of $T_N$. 
Takubo {\it et.al} estimated the amplitude of $\pi$ hopping is half of $t$ \cite{takubo2006}.

In summary, we obtained the observed uudd-MO without OO for the intermediate-coupling regime of the three-band Hubbard model by using QMC simulations 
complemented by thermal and quantum annealing techniques. Preliminary results for $t=0.3$ (eV) (not shown in this paper) are also confirming the existence of the OO proposed by  Tsunetsugu and Motome \cite{tsunetsugu2003}. 
A complete the phase diagram that includes to the transition to the strong-coupling regime will be presented in a future work.

\section*{Acknowledgement}
I appreciate G-W. Chern, N. Parkins, C.D. Batista for fruitful discussions. 
Work at the LANL was performed under the auspices of the
U.S.\ DOE contract No.~DE-AC52-06NA25396 through the LDRD program.
The numerical simulations in this paper were implemented at the National Energy Research Scientific Computing Center.


\bibliographystyle{elsarticle-num}
\bibliography{kato}







\end{document}